\documentclass[11pt,a4paper]{article}

\usepackage{amsfonts,amsmath,amssymb,bm}

\begin{document}

\title{\bf Quantization of massive Weyl fields in vacuum}
\author{Maxim~Dvornikov$^{a,b}$\footnote{{\bf e-mail}: maxim.dvornikov@usp.br}
\\
$^a$ \small{\em Institute of Physics, University of S\~{a}o Paulo} \\
\small{\em CP 66318, CEP 05315-970 S\~{a}o Paulo, SP, Brazil}
\\
$^b$ \small{\em Pushkov Institute of Terrestrial Magnetism, Ionosphere} \\
\small{\em and Radiowave Propagation (IZMIRAN)} \\
\small{\em 142190 Moscow, Troitsk, Russia}
}
\date{}
\maketitle

\begin{abstract}
We briefly review the main methods for the description of massive Weyl fields in vacuum. On the classical level we discuss Weyl fields expressed through Grassmann variables as well as having spinors with commuting components. In both approaches we quantize the system. We get the correct anticommutation relations between creation and annihilation operators, which result in the proper form of the total energy of the field. However, the commuting classical Weyl fields require the new method of quantization.
\end{abstract}

\section{Introduction}

The studies of Majorana fermions play an important role in various branches of modern physics, such as high energy physics and superconductivity. For the first time the consistent treatment of massive Majorana fermions were presented in Ref.~\cite{Cas57}. Then a system of coupled Majorana fermions was studies in Ref.~\cite{SchVal81}, in connection to the problem of massive neutrinos propagation in external electromagnetic fields. Recently it was proposed~\cite{Kit01} that the quasiparticles having Majorana properties can be excited in a superconductor.

It was claimed in Ref.~\cite{SchVal81}, that the dynamics of a Majorana fermion on the classical level must be formulated using anticommuting  Grassmann variables ($g$-numbers). We remind that in case of a classical Dirac fermion one can use both commuting $c$-numbers or $g$-numbers to describe its dynamics. To study this problem in details in Ref.~\cite{Ahl11} the new fermionic field, which belongs to a non-standard Wigner class, was introduced. It was also shown that such an object has some properties of a $c$-number Majorana field. The application of the results of Ref.~\cite{Ahl11} to the dark matter problem was studied in Ref.~\cite{RocBerSil11}.

This work is a brief review of the main methods for the description of a massive Majorana particle in vacuum. It is organized in the following way. In Sec.~\ref{MAJWEY} we discuss the different representations of fermionic particles which are equivalent to their antiparticles. The $g$-number treatment of classical Weyl fields is presented in Sec.~\ref{sec:CLASS}. Then, in Sec.~\ref{sec:QUANT}, we carry out a canonical quantization of a massive Weyl field. In Sec.~\ref{CLASSICAL} we propose an alternative way for the description of massive Weyl fields, which are represented in terms of $c$-number spinors. The new method of quantization of $c$-number Weyl fields is formulated in Sec.~\ref{QUANTUM}. Finally, in Sec.~\ref{CONCL}, we briefly summarize out results.

\section{Majorana and Weyl fermions\label{MAJWEY}}

In this section we shall present two alternative approaches for the description of $1/2$-spin fermion fields with identical particle and antiparticle degrees of freedom. We shall concentrate on Majorana and Weyl representations for such fields.

A fermionic $1/2$-spin field $\psi$ obeys the Dirac equation,
\begin{equation}\label{eq:Direq}
  (\mathrm{i}\gamma^{\mu}\partial_{\mu}-m)\psi=0,
\end{equation}
where $m$ is the mass of the field. The Dirac $\gamma$-matrices
in the chiral representation can be chosen as~\cite{ItzZub80}
\begin{equation}
  \gamma^{\mu} =
  \left(
    \begin{array}{cc}
      0 & -\sigma^{\mu} \\
      -\tilde{\sigma}^{\mu} & 0
    \end{array}
  \right),
\end{equation}
where $\sigma^{\mu}=(\sigma^{0},-\bm{\sigma})$, $\tilde{\sigma}^{\mu}=(\sigma^{0},\bm{\sigma})$,
$\sigma^{0}=\sigma_{0}$ is the unit $2\times2$ matrix, and $\bm{\sigma}$
are the Pauli matrices.

We shall suppose that the four component spinor $\psi$ satisfies the Majorana condition in an extended sense~\cite[see p.~277]{FukYan03},
\begin{equation}\label{Majdef}
  \mathcal{C}\bar{\psi}^\mathrm{T} = \kappa \psi,
  \quad
  \mathcal{C} = \mathrm{i} \gamma^2 \gamma^0,
\end{equation}
where $\kappa$ is complex number with unit absolute value. In the following we shall suppose that $\kappa = 1$. A fermionic field satisfying condition~\eqref{Majdef} is called a Majorana field. It is demonstrated in Ref.~\cite[see pp.~80 and~99]{BerLifPit82} the one can choose a special representation of spinors and $\gamma$-matrices, called the Majorana representation, in which the Majorana condition~\eqref{Majdef} is equivalent to the complex conjugation, i.e. Majorana spinors are real.

It is, however, more convenient to deal with a two component Weyl spinor, $\eta$, which can be introduced as
\begin{equation}\label{eq:Majspinor}
  \psi =
  \left(
    \begin{array}{c}
      \mathrm{i}\sigma_{2}\eta^{*} \\
      \eta
    \end{array}
  \right).
\end{equation}
Note that this spinor satisfies the Majorana condition~\eqref{Majdef}. We can get a wave equation for $\eta$~\cite[see pp.~292--296]{FukYan03},
\begin{equation}\label{eq:Weyleq}
  \sigma^{\mu}\partial_{\mu}\eta+m\sigma_{2}\eta^{*}=0,
\end{equation}
using the more general Dirac equation~(\ref{eq:Direq}).

\section{Classical massive Weyl field in vacuum: $g$-numbers\label{sec:CLASS}}

In this section we shall describe the variational procedure, based on the Lagrange formalism, to get the wave equation for a classical massive Weyl field in vacuum. Then we apply the canonical formalism to reproduce the dynamics of the system. In the present section we suppose that the spinor $\eta$ has anticommuting $g$-number components~\cite{DvoGit12}.

The wave equation~(\ref{eq:Weyleq}) can be obtained using the standard
variational procedure, $\partial_{r}\mathcal{L}/\partial\eta^{*}=0$,
where $\partial_{r}/\partial\eta^{*}$ is the right derivative~\cite{BerMar77},
on the basis of the following Lagrangian:
\begin{equation}\label{eq:WeylLagr}
  \mathcal{L} =\mathrm
  {i} \eta^{\dagger} \sigma^{\mu} \partial_{\mu} \eta - \frac{\mathrm{i}}{2} m
  \left(
    \eta^{\mathrm{T}} \sigma_{2} \eta - \eta^{\dagger} \sigma_{2} \eta^{*}
  \right).
\end{equation}
It should be noted that within the Lagrange formalism it is crucial to deal with $g$-number wave functions.
Otherwise the mass term in Eq.~(\ref{eq:WeylLagr}) is vanishing.

On the basis of Eq.~(\ref{eq:WeylLagr}) we can find the canonical
momenta as
\begin{equation}\label{eq:canmom}
 \pi =
 \frac{\partial_{r}\mathcal{L}}{\partial\dot{\eta}} = \mathrm{i}\eta^{*},
 \quad
 \pi^{*} = \frac{\partial_{r}\mathcal{L}}{\partial\dot{\eta}^{*}} = 0.
\end{equation}
Eq.~(\ref{eq:canmom}) means that the system in question has two
primary second class constraints,
\begin{equation}\label{eq:constr}
 \Phi_{1} = \pi-\mathrm{i}\eta^{*} = 0,
 \quad
 \Phi_{2} = \pi^{*} = 0.
\end{equation}
Thus besides the Hamiltonian,
\begin{equation}\label{eq:Ham}
  \mathcal{H} = \pi\dot{\eta}+\pi^{*}\dot{\eta}^{*} - \mathcal{L} =
  \mathrm{i}\eta^{\dagger}(\bm{\sigma}\nabla)\eta + \frac{\mathrm{i}}{2}m
  \left(
    \eta^{\mathrm{T}}\sigma_{2}\eta - \eta^{\dagger}\sigma_{2}\eta^{*}
  \right),
\end{equation}
we have to consider the extended Hamiltonian,
\begin{equation}\label{eq:extHam}
  \mathcal{H}_{1} = \mathcal{H}+\Phi_{1}\lambda_{1} + \lambda_{2}\Phi_{2},
\end{equation}
which account for the constraints. Here $\lambda_{1,2}$ are the Lagrange
multipliers which are odd $g$-numbers.

To exclude the parameters $\lambda_{1,2}$, we should take into account
the constraints conservation,
\begin{equation}\label{eq:constcons}
  \left\{
    \Phi_{1},\mathcal{H}_{1}
  \right\} =
  \left\{
    \Phi_{2},\mathcal{H}_{1}
  \right\} = 0,
\end{equation}
where~\cite[see p.~76]{GitTyt90}
\begin{equation}\label{eq:Poisbrac}
  \left\{
    F,G
  \right\} =
  \frac{\partial_{r}F}{\partial\eta} \frac{\partial_{l}G}{\partial\pi} -
  (-1)^{P_{F}P_{G}} \frac{\partial_{r}G}{\partial\eta} \frac{\partial_{l}F}{\partial\pi} +
  \left(
    \eta \to \eta^{*}, \pi \to \pi^{*}
  \right),
\end{equation}
is the Poisson bracket of two dynamical variables $F$ and $G$, $\partial_{l}/\partial\eta$
is the left derivative~\cite{BerMar77}, and $P_{F}$ is the Grassmann parity of the
function $F$. Using Eqs.~(\ref{eq:constr})-(\ref{eq:Poisbrac})
we get $\lambda_{1,2}$ as
\begin{equation}
  \lambda_{1} =
  (\bm{\sigma}\nabla)\eta - m\sigma_{2}\eta^{*},
  \quad
  \lambda_{2} =
  -
  \left(
    \nabla\eta^{\dagger}
  \right)
  \bm{\sigma} + m\eta^{\mathrm{T}}\sigma_{2}.
\end{equation}

The presence of the second class constraints in a system implies the
replacement of the Poisson brackets by the Dirac brackets defined
as~\cite{Dir01}
\begin{equation}\label{eq:Dirbrac}
  \left\{
    F,G
  \right\}_{\mathrm{D}} =
  \left\{
    F,G
  \right\} -
  \left\{
    F,\Phi_{l}
  \right\}
  C_{ls}
  \left\{
    \Phi_{s},G
  \right\},
\end{equation}
where the matrix $(C_{ls})$ has the components $C_{ls}=\left\{ \Phi_{l},\Phi_{s}\right\} ^{-1}$.
Using Eqs.~(\ref{eq:constr}) and~(\ref{eq:Poisbrac}) we get this
matrix in the explicit form,
\begin{equation}\label{eq:Cdef}
  C =
  \begin{pmatrix}
    0 & \mathrm{i}\\
    \mathrm{i} & 0
  \end{pmatrix}
  \delta^{3}(\mathbf{x}-\mathbf{y}).
\end{equation}
On the basis of Eqs.~(\ref{eq:Dirbrac}) and~(\ref{eq:Cdef}) we
can obtain the nonzero Dirac brackets
\begin{align}
  \left\{
    \eta(\mathbf{x},t),\eta^{*}(\mathbf{y},t)
  \right\}_{\mathrm{D}} = &
  -\mathrm{i}\delta^{3}(\mathbf{x}-\mathbf{y}),\\
  \left\{
    \eta(\mathbf{x},t),\pi(\mathbf{y},t)
  \right\}_{\mathrm{D}} = &
  \delta^{3}(\mathbf{x}-\mathbf{y}).
  \label{eq:fundDirbrac}
\end{align}
Note that the expression of the fundamental Dirac bracket~(\ref{eq:fundDirbrac})
will be later used in the quantization of the system.

Using Eqs.~\eqref{eq:Ham} and~\eqref{eq:Dirbrac} we can also reproduce the full set of wave equations,
\begin{align}\label{eq:Weyleqfull}
  \dot{\eta}= &
  \left\{
    \eta, \mathcal{H}
  \right\}_{\mathrm{D}} =
  (\bm{\sigma}\nabla)\eta - m\sigma_{2}\eta^{*},\nonumber \\
  \dot{\eta}^{*}= &
  \left\{
    \eta^{*},\mathcal{H}
  \right\}_{\mathrm{D}} =
  \left(
    \bm{\sigma}^{\mathrm{T}}\nabla
  \right)
  \eta^{*} + m\sigma_{2}\eta,\nonumber \\
  \dot{\pi}= &
  \left\{
    \pi,\mathcal{H}
  \right\}_{\mathrm{D}} =
  \mathrm{i}
  \left(
    \bm{\sigma}^{\mathrm{T}}\nabla
  \right)
  \eta^{*} + \mathrm{i}m\sigma_{2}\eta,\nonumber \\
  \dot{\pi}^{*} = &
  \left\{
    \pi^{*},\mathcal{H}
  \right\}_{\mathrm{D}} = 0.
\end{align}
We can see in Eq.~(\ref{eq:Weyleqfull}) that the evolution equation
for $\eta$ coincides with Eq.~(\ref{eq:Weyleq}) derived directly
from the Dirac equation~(\ref{eq:Direq}). Using Eq.~(\ref{eq:constr})
it is convenient to rewrite the wave equations for $\eta$ and $\pi$,
\begin{align}\label{eq:Weyleqetapi}
  \dot{\eta} & =
  (\bm{\sigma}\nabla)\eta + \mathrm{i}\sigma_{2}m\pi,\nonumber \\
  \dot{\pi} & =
  \left(
    \bm{\sigma}^{\mathrm{T}}\nabla
  \right)\pi + \mathrm{i}\sigma_{2}m\eta,
\end{align}
to exclude the operation of the complex conjugation.

\section{Canonical quantization\label{sec:QUANT}}

In this section we carry out a canonical quantization of a massive
Weyl field in vacuum. We use the basic results obtained in Sec.~\ref{sec:CLASS} and replace the Dirac brackets of canonical variables with the anticommutators. Then we introduce the creation and annihilation operators and establish the anticommutation relations for them.

The two component Weyl spinor $\eta$, described in Sec.~\ref{sec:CLASS},
corresponds to $1/2$-spin particles, i.e. it is fermionic. For the
quantization of such a system one should (i) replace the $g$-number
classical wave functions with operators $\eta\to\hat{\eta}$
and $\pi\to\hat{\pi}$; and (ii) define the equal time anticommutator
for these operators as~\cite[see pp.~81--86]{GitTyt90}
\begin{equation}\label{eq:canquantfundanticom}
  \left[
    \hat{\eta}(\mathbf{x},t), \hat{\pi}(\mathbf{y},t)
  \right]_{+} =
  \mathrm{i}
  \left.
    \left\{
      \eta(\mathbf{x},t),\pi(\mathbf{y},t)
    \right\}_{\mathrm{D}}
  \right|_{\eta=\hat{\eta},\pi=\hat{\pi}} = \mathrm{i}\delta^{3}(\mathbf{x}-\mathbf{y}),
\end{equation}
where the fundamental Dirac bracket is given in Eq.~(\ref{eq:fundDirbrac}).

Using Eq.~(\ref{eq:constr}) it is convenient to rewrite Eq.~(\ref{eq:canquantfundanticom})
as
\begin{equation}\label{eq:anticommod}
  \left[
    \hat{\eta}(t,\mathbf{x}),\hat{\eta}^{\dagger}(t,\mathbf{y})
  \right]_{+} =
  \delta^{3}(\mathbf{x}-\mathbf{y}).
\end{equation}
Now we can find the general solution of Eq.~(\ref{eq:Weyleq}) in
the following form~\cite[see pp.~292--296]{FukYan03}:
\begin{align}\label{eq:solWeyleq}
  \hat{\eta}(x) = &
  \int \frac{\mathrm{d}^{3}\mathbf{p}}{(2\pi)^{3/2}} \sqrt{\frac{E+|\mathbf{p}|}{2E}}
  \notag
  \\
  & \times
  \left[
    \left(
      \hat{a}_{-}w_{-}-\frac{m}{E+|\mathbf{p}|}\hat{a}_{+}w_{+}
    \right)
    e^{-\mathrm{i}px}
    +
    \left(
      \hat{a}_{+}^{\dagger}w_{-}+\frac{m}{E+|\mathbf{p}|}\hat{a}_{-}^{\dagger}w_{+}
    \right)
    e^{\mathrm{i}px}
  \right],
\end{align}
where $p^\mu = (E,\mathbf{p})$, $E=\sqrt{|\mathbf{p}|^{2}+m^{2}}$, $\hat{a}_{\pm}^{\dagger}(\mathbf{p})$
and $\hat{a}_{\pm}(\mathbf{p})$ are the creation and annihilation
operators, and
\begin{equation}
  w_{+}=
  \left(
    \begin{array}{c}
      e^{-\mathrm{i}\phi/2}\cos\theta/2\\
      e^{\mathrm{i}\phi/2}\sin\theta/2
    \end{array}
  \right),
  \quad
  w_{-}=
  \left(
    \begin{array}{c}
      -e^{-\mathrm{i}\phi/2}\sin\theta/2\\
      e^{\mathrm{i}\phi/2}\cos\theta/2
    \end{array}
  \right),
\end{equation}
are the chiral amplitudes~\cite[see p.~86]{BerLifPit82}. Here the angles $\phi$ and $\theta$ fix
the direction of the particle momentum, $\mathbf{p}=|\mathbf{p}|(\cos\phi\sin\theta,\sin\phi\sin\theta,\cos\theta)$.

On the basis of Eq.~(\ref{eq:solWeyleq}), we can find that Eq.~(\ref{eq:anticommod})
is satisfied, provided that the canonical anticommutation relations,
\begin{equation}\label{eq:cananticomaad}
  \left[
    \hat{a}_{\sigma}(\mathbf{k}), \hat{a}_{\sigma'}^{\dagger}(\mathbf{k}')
  \right]_{+} =
  \delta_{\sigma\sigma'}\delta^{3}(\mathbf{k}-\mathbf{k}'),
  \quad
  \left[
    \hat{a}_{\sigma}(\mathbf{k}), \hat{a}_{\sigma'}(\mathbf{k}')
  \right]_{+} = 0,
  \quad
  \left[
    \hat{a}_{\sigma}^{\dagger}(\mathbf{k}), \hat{a}_{\sigma'}^{\dagger}(\mathbf{k}')
  \right]_{+}=0,
\end{equation}
are valid for the operators $\hat{a}_{\sigma}(\mathbf{p})$ and $\hat{a}_{\sigma}^{\dagger}(\mathbf{p})$.

To verify the validity of the expansion~(\ref{eq:solWeyleq}) we
can use it for the calculation of the total energy of a massive Weyl
field. Using Eqs.~(\ref{eq:Ham}), (\ref{eq:solWeyleq}), and~(\ref{eq:cananticomaad})
we get for the total energy,
\begin{equation}
  E_{\mathrm{tot}} =
  \int \mathrm{d}^{3}\mathbf{r} \mathcal{H} =
  \int \mathrm{d}^{3}\mathbf{p} E
  \left(
    \hat{a}_{-}^{\dagger}\hat{a}_{-} + \hat{a}_{+}^{\dagger}\hat{a}_{+}
  \right) +
  \text{divergent terms},
\end{equation}
where ``divergent terms'' contain $\delta^{3}(0)$ and can be removed
by the normal ordering of operators.

\section{Classical massive Weyl field in vacuum: $c$-numbers\label{CLASSICAL}}

As we mentioned in Sec.~\ref{sec:CLASS} the mass term in Eq.~\eqref{eq:WeylLagr} vanishes if the spinor $\eta$ has $c$-number components. It might be an indication that $g$-number description is essential for a massive Weyl field even on a classical level. It is, however, known that, if one works with classical fermionic fields, $c$- and $g$-numbers approaches are equivalent~\cite{Wei96}.

To clarify this issue we remind that the Lagrange formalism is not a unique way for the treatment of a dynamical system. Following Ref.~\cite{Dvo12FP}, we can consider the Hamilton formalism for the description of a massive Weyl field represented by $c$-number spinors. Let us consider the following Hamiltonian:
\begin{equation}\label{Hamclass}
  H[\eta,\eta^{*{}},\pi,\pi^{*{}}] = \int \mathrm{d}^3\mathbf{r}
  \big\{
    \pi^\mathrm{T} (\bm{\sigma}\nabla) \eta -
    (\eta^{*{}})^\mathrm{T} (\bm{\sigma}\nabla) \pi^{*{}}
    +
    m
    \left[
      (\eta^{*{}})^\mathrm{T} \sigma_2 \pi +
      (\pi^{*{}})^\mathrm{T} \sigma_2 \eta
    \right]
  \big\},
\end{equation}
which is a functional of independent canonical variables $(\eta,\pi)$ and $(\eta^{*{}},\pi^{*{}})$. We can see that $H$ is real.

Using the classical field theory version of the canonical Hamilton
equation,
\begin{equation}\label{etaclass}
  \dot{\eta}  = \frac{\delta H}{\delta \pi} =
  (\bm{\sigma}\nabla)\eta - m \sigma_2 \eta^{*{}},
\end{equation}
we obtain Eq.~\eqref{eq:Weyleq} for a massive Weyl field. As it was shown in Ref.~\cite{Dvo12FP}, the canonical variables $\eta$ and $\pi$ evolve independently. Thus we cannot find the relation between
the canonical momenta and the ``velocities", $\dot{\eta}$ and
$\dot{\eta}^{*{}}$, to construct a Lagrangian~\cite{Gan75}.

Despite that no conventional Lagrange formalism can be applied for the description of our system, we can construct an extended Lagrangian, $\tilde{\mathcal{L}}$, which also includes the momenta, $\pi$ and  $\pi^{*{}}$, as well as their time derivatives, $\dot{\pi}$ and $\dot{\pi}^{*{}}$, as independent variables. Let us choose the extended Lagrangian as~\cite{FadJac88}
\begin{equation}\label{extLagr}
  \tilde{\mathcal{L}} =
  \pi^\mathrm{T} \dot{\eta} + (\pi^{*{}})^\mathrm{T} \dot{\eta}^{*{}} -
  \left[
    \pi^\mathrm{T} (\bm{\sigma}\nabla) \eta -
    (\eta^{*{}})^\mathrm{T} (\bm{\sigma}\nabla) \pi^{*{}}
  \right]
  -
  m
  \left[
    (\eta^{*{}})^\mathrm{T} \sigma_2 \pi +
    (\pi^{*{}})^\mathrm{T} \sigma_2 \eta
  \right].
\end{equation}
Here we use the opportunity to correct the mistake in Eq.~(14) in Ref.~\cite{Dvo12FP}. There is an extra erroneous factor $1/2$ in the spatial derivatives term in the expression for the extended Lagrangian.

Varying this Lagrangian with respect to $\pi$ one can reproduce Eq.~\eqref{eq:Weyleq}. Again we can see that, in frames of the modified Lagrange formalism, two groups of variables, $(\eta,\eta^{*{}})$ and $(\pi,\pi^{*{}})$, evolve independently. One can say that the evolution of the system, based on Eq.~\eqref{extLagr}, is an extended Lagrange dynamics in the analogy to the extended Hamilton formalism~\cite[see pp.~13--21]{GitTyt90}.

\section{New method for the quantization of a massive Weyl field\label{QUANTUM}}

In this section we carry out a quantization of a classical massive Weyl field described by $c$-number spinors. In particular, we show that there are two independent ways of the quantization.

Analogously to Sec.~\ref{sec:QUANT} we find the plane wave representation for $\eta$ and $\xi = \mathrm{i} \sigma_2 \pi$ as
\begin{align}\label{etaxisol}
  \eta(x) = & \frac{1}{2}
  \int \frac{\mathrm{d}^3\mathbf{p}}{(2\pi)^{3/2}}
  \sqrt{1 + \frac{E}{|\mathbf{p}|}}
  \notag
  \\
  & \times
  \left[
    \left(
      \hat{a}_{-{}} w_{-{}} -
      \frac{m}{E+|\mathbf{p}|} \hat{a}_{+{}} w_{+{}}
    \right) e^{-\mathrm{i}px}
    +
    \left(
      \hat{a}_{+{}}^\dagger w_{-{}} +
      \frac{m}{E+|\mathbf{p}|} \hat{a}_{-{}}^\dagger w_{+{}}
    \right) e^{\mathrm{i}px}
  \right],
  \notag
  \displaybreak[2]
  \\
  \xi(x) = & \frac{\mathrm{i}}{2}
  \int \frac{\mathrm{d}^3\mathbf{p}}{(2\pi)^{3/2}}
  \sqrt{1 + \frac{E}{|\mathbf{p}|}}
  \notag
  \\
  & \times
  \left[
    \left(
      \hat{b}_{+{}} w_{+{}} +
      \frac{m}{E+|\mathbf{p}|} \hat{b}_{-{}} w_{-{}}
    \right) e^{-\mathrm{i}px}
    +
    \left(
      \hat{b}_{-{}}^\dagger w_{+{}} -
      \frac{m}{E+|\mathbf{p}|} \hat{b}_{+{}}^\dagger w_{-{}}
    \right) e^{\mathrm{i}px}
  \right],
\end{align}
where the helicity amplitudes, $w_\sigma$, were defined in Sec.~\ref{sec:QUANT}. Note that the expansion coefficients, $\hat{a}_{\pm{}}$ and $\hat{b}_{\pm{}}$, which are supposed to be operators, are different for $\eta$ and $\xi$ since we showed in Sec.~\ref{CLASSICAL} that these fields evolve independently.

On the basis of Eqs.~\eqref{Hamclass} and \eqref{etaxisol}, after a bit lengthy but straightforward
calculations, we get the Hamiltonian expressed in terms of the
operators $a_{\pm{}}(\mathbf{p})$ and $b_{\pm{}}(\mathbf{p})$ and
their conjugate,
\begin{align}\label{Hamquant}
  H = & \frac{1}{4} \int \mathrm{d}^3\mathbf{p}
  E
  \left(
    1 + \frac{E}{|\mathbf{p}|}
  \right)
  \bigg\{
  \Big\{
    \hat{a}_{-{}}^\dagger(\mathbf{p}) \hat{b}_{-{}}(\mathbf{p}) +
    \hat{b}_{-{}}^\dagger(\mathbf{p}) \hat{a}_{-{}}(\mathbf{p}) -
    \hat{a}_{+{}}(\mathbf{p}) \hat{b}_{+{}}^\dagger(\mathbf{p}) -
    \hat{b}_{+{}}(\mathbf{p}) \hat{a}_{+{}}^\dagger(\mathbf{p})
    \notag
    \\
    & +
    \left(
      \frac{m}{E+|\mathbf{p}|}
    \right)^2
    \left[
      \hat{a}_{-{}}(\mathbf{p}) \hat{b}_{-{}}^\dagger(\mathbf{p}) +
      \hat{b}_{-{}}(\mathbf{p}) \hat{a}_{-{}}^\dagger(\mathbf{p}) -
      \hat{a}_{+{}}^\dagger(\mathbf{p}) \hat{b}_{+{}}(\mathbf{p}) -
      \hat{b}_{+{}}^\dagger(\mathbf{p}) \hat{a}_{+{}}(\mathbf{p})
    \right]
    \Big\}
    \notag
    \displaybreak[2]
    \\
    & +
    \mathrm{i}
    \frac{m}{|\mathbf{p}|}
    \Big\{
      e^{-2\mathrm{i}Et}
      \left[
        \hat{a}_{-{}}(\mathbf{p}) \hat{b}_{-{}}(-\mathbf{p}) +
        \hat{b}_{-{}}(-\mathbf{p}) \hat{a}_{-{}}(\mathbf{p}) +
        \hat{b}_{+{}}(-\mathbf{p}) \hat{a}_{+{}}(\mathbf{p}) +
        \hat{a}_{+{}}(\mathbf{p}) \hat{b}_{+{}}(-\mathbf{p})
      \right]
      \notag
      \\
      & +
      e^{2\mathrm{i}Et}
      \left[
        \hat{a}_{-{}}^\dagger(\mathbf{p}) \hat{b}_{-{}}^\dagger(-\mathbf{p}) +
        \hat{b}_{-{}}^\dagger(-\mathbf{p}) \hat{a}_{-{}}^\dagger(\mathbf{p}) +
        \hat{b}_{+{}}^\dagger(-\mathbf{p}) \hat{a}_{+{}}^\dagger(\mathbf{p}) +
        \hat{a}_{+{}}^\dagger(\mathbf{p}) \hat{b}_{+{}}^\dagger(-\mathbf{p})
      \right]
    \Big\}
  \bigg\}.
\end{align}
Now we establish the following relation between the independent
operators $\hat{a}_{\pm{}}(\mathbf{p})$ and $\hat{b}_{\pm{}}(\mathbf{p})$:
\begin{equation}\label{Majcondoper}
  \hat{a}_{\pm{}}(\mathbf{p}) = \hat{b}_{\pm{}}(\mathbf{p}),
\end{equation}
and the analogous expression for the conjugate operators. We will
choose the operators $\hat{a}_{\pm{}}(\mathbf{p})$ as the basic ones and
assume that they obey the anticommutation relations,
\begin{equation}\label{anticomrel}
  \left[
    \hat{a}_\sigma(\mathbf{k}),\hat{a}_{\sigma'}^\dagger(\mathbf{p})
  \right]_{+{}} =
  \delta_{\sigma\sigma'} \delta^3(\mathbf{k} - \mathbf{p}),
\end{equation}
with all the other anticommutators being equal to zero. In this
case the time dependent terms in Eq.~\eqref{Hamquant} are washed
out. Using Eqs.~\eqref{Majcondoper} and~\eqref{anticomrel} we can
recast Eq.~\eqref{Hamquant} into the form
\begin{equation}\label{totenquant}
  H = \int \mathrm{d}^3\mathbf{p}
  \thinspace E
  (\hat{a}_{-{}}^\dagger \hat{a}_{-{}} +
  \hat{a}_{+{}}^\dagger \hat{a}_{+{}}) + \text{divergent terms},
\end{equation}
which shows that the total energy of a Weyl field is a sum of the
energies of elementary oscillators corresponding to the negative
and the positive helicity states.

In the canonical formalism the total momentum of our system can be
calculated using the expression,
\begin{equation}
  \mathbf{P} = \int \mathrm{d}^3\mathbf{r}
  \left[
    (\eta^{*{}})^\mathrm{T} \nabla \pi^{*{}} -
    \pi^\mathrm{T} \nabla \eta
  \right],
\end{equation}
which is obtained by the spatial integration of the $T^{i0}$
component of the energy-momentum tensor $T^{\mu\nu}$. Omitting the
detailed calculations and with help of Eqs.~\eqref{etaxisol} \eqref{Majcondoper}, and~\eqref{anticomrel} we get
the following formula for the quantized momentum of the Weyl
field:
\begin{equation}\label{totmomquant}
  \mathbf{P} = \int \mathrm{d}^3\mathbf{p}
  \thinspace
  \mathbf{p} (\hat{a}_{-{}}^\dagger \hat{a}_{-{}} +
  \hat{a}_{+{}}^\dagger \hat{a}_{+{}}) + \text{divergent terms},
\end{equation}
which has the analogous structure as Eq.~\eqref{totenquant}.

We can also show that besides the aforementioned way to establish the correct anticommutation relation for the operators $\hat{a}_{\pm{}}$, cf. Eq.~\eqref{anticomrel}, there is another way to quantize a massive Weyl field. Instead
of Eq.~\eqref{Majcondoper} we may choose the following relation
between the operators:
\begin{equation}\label{Majcondoperalt}
  \hat{a}_{\pm{}}(\mathbf{p}) = \hat{b}_{\mp{}}(\mathbf{p}),
\end{equation}
with the condition~\eqref{anticomrel} still being held true for
the operators $\hat{a}_{\pm{}}$. As it was shown in Ref.~\cite{Dvo12FP}, in this case we can also get the correct form of the total energy and momentum.

\section{Conclusion\label{CONCL}}

In the present work we have discussed several important issues in the description of massive Weyl fields in vacuum. Firstly, in Sec.~\ref{sec:CLASS} we applied a formalism of $g$-numbers for the treatment of a single massive Weyl field. It was demonstrated that there are second class constraints in this system. We have reproduced the wave equations using the calculated Dirac brackets. Then, in Sec.~\ref{sec:QUANT} we carried out a canonical quantization of the system.

It was mentioned in Sec.~\ref{CLASSICAL} that, in frames of the Lagrange formalism, it is crucial that a massive Weyl field is described by $g$-numbers even on the classical level. Otherwise the mass term in Eq.~\eqref{eq:WeylLagr} is vanishing. To overcome this difficulty in the classical theory of massive Weyl fields, in Sec.~\ref{CLASSICAL} we developed an approach based on the Hamilton formalism for the description of Weyl spinors which have $c$-number components. In Sec.~\ref{CLASSICAL} we have also shown that one can apply an extended Lagrange dynamics for the treatment of this system. Finally, in Sec.~\ref{QUANTUM} we have proposed the new method of quantization of a classical Weyl field described in Sec.~\ref{CLASSICAL}. Note that the detailed description of the methods used in Secs.~\ref{QUANTUM} and~\ref{CLASSICAL} is provided in Refs.~\cite{Dvo12FP,Dvo12NPB}.

The results of the present work are of great importance for the modern particle physics since neutrinos, which are experimentally proven to be massive particles (see, e.g., Ref~\cite{nuosc}), are the most prominent candidates to be described in terms of Weyl fields~\cite{Kob80}. Moreover, effective Majorana fields in one and two dimensions can be encountered in the effective theory of $p$-waves superconductors~\cite{LeiFle12}.

\section*{Acknowledgments}

I am very thankful to D.~M.~Gitman for helpful discussions and to FAPESP (Brazil) for a grant.


\begin{thebibliography}{40}

\bibitem{Cas57}
  K.~M.~Case,
  Phys. Rev. \textbf{107}, 307 (1957).

\bibitem{SchVal81}
  J.~Schechter and J.~W.~F.~Valle,
  Phys. Rev. D \textbf{24}, 1883 (1981).

\bibitem{Kit01}
  A.~Yu.~Kitaev,
  Phys. Usp. (Suppl.) \textbf{44}, 131 (2001),
  cond-mat/0010440.

\bibitem{Ahl11}
  D.~V.~Ahluwalia, C.-Y.~Lee, and D.~Schritt,
  Phys. Rev. D \textbf{83}, 065017 (2011),
  arXiv:0911.2947~[hep-ph].

\bibitem{RocBerSil11}
  R.~da~Rocha, A.~E.~Bernardini, and J.~M.~Hoff~da~Silva,
  JHEP \textbf{1104}, 110 (2011),
  arXiv:0911.2947~[hep-ph].

\bibitem{ItzZub80}
  C.~Itzykson and J.-B.~Zuber,
  \textit{Quantum Field Theory}
  (McGraw-Hill, New York, 1980), p.~694.

\bibitem{FukYan03}
  M.~Fukugita and T.~Yanagida,
  \textit{Physics of Neutrinos and Applications to Astrophysics}
  (Springer, Berlin, 2003).

\bibitem{BerLifPit82}
  V.~B.~Berestetski\u{\i}, E.~M.~Lifshitz, and L.~P.~Pitaevski\u{\i},
  \textit{Quantum Electrodynamics}
  (Pergamon, Oxford, 1980), 2nd ed.

\bibitem{DvoGit12}
  M.~Dvornikov and D.~M.~Gitman,
  Phys. Rev. D \textbf{87}, 025027 (2013),
  arXiv:1211.5367~[hep-th].

\bibitem{BerMar77}
  F.~A.~Berezin and M.~S.~Marinov,
  Ann. Phys. (N.Y.) \textbf{104}, 336 (1977).

\bibitem{GitTyt90}
  D.~M.~Gitman and I.~V.~Tyutin,
  \textit{Quantization of Fields with Constraints}
  (Springer, Berlin, 1990).

\bibitem{Dir01}
  P.~A.~M.~Dirac,
  \textit{Lectures on Quantum Mechanics}
  (Dover, New York, 2001), p.~41.

\bibitem{Wei96}
  S. Weinberg,
  \textit{The Quantum Theory of Felds: Foundations}
  (Cambridge University Press, Cambridge, 1996),
  2nd ed., pp.~292--338 and pp.~399--413.

\bibitem{Dvo12FP}
  M.~Dvornikov,
  Found. Phys. \textbf{42}, 1469 (2012),
  arXiv:1106.3303~[hep-th].

\bibitem{Gan75}
  F.~Gantmacher,
  \textit{Lectures in Analytical Mechanics}
  (Mir Publishers, Moscow, 1975), pp.~71--80.

\bibitem{FadJac88}
  L.~Faddeev and R.~Jackiw,
  Phys. Rev. Lett. \textbf{60}, 1692 (1988).

\bibitem{Dvo12NPB}
  M.~Dvornikov,
  Nucl. Phys. B \textbf{855}, 760 (2012),
  arXiv:1108.5043~[hep-ph].

\bibitem{nuosc}
  F.~P.~An \textit{et al.} (Daya Bay Collaboration),
  Phys. Rev. Lett. \textbf{108}, 171803 (2012),
  arXiv:1203.1669~[hep-ex];
  Y.~Abe \textit{et al.} (Double Chooz Collaboration),
  Phys. Rev. Lett. \textbf{108}, 131801 (2012),
  arXiv:1112.6353~[hep-ex].

\bibitem{Kob80}
  I.~Yu.~Kobzarev \textit{et al.},
  Sov. J. Nucl. Phys. \textbf{32}, 823 (1980).

\bibitem{LeiFle12}
  M.~Leijnse and K.~Flensberg,
  Semicond. Sci. Technol. \textbf{27}, 124003 (2012),
  arXiv:1206.1736~[cond-mat.mes-hall].

\end{thebibliography}
\end{document}